\documentstyle[preprint,tighten,aps]{revtex}
\begin{document}
\preprint{$\strut\hbox to 5 truecm{\hfil INFNNA-IV-93-31}
\atop
\strut\hbox to 5 truecm{\hfil UTS-DFT-93-29}$}
\date{November 1993}
\title{Kaon decay interferometry\\
as meson dynamics probes\footnote{
Work supported in part by the Human Capital and Mobility Program, EEC
Contract N. CHRX-CT920026.}}
\author{G. D'Ambrosio}
\address{Istituto Nazionale di Fisica Nucleare, Sezione di Napoli,
I-80125 Italy\\
Dipartimento di Scienze Fisiche, Universit\`{a} di Napoli, I-80125 Italy}
\author{N. Paver}
\address{Dipartimento di Fisica Teorica, Universit\`{a} di Trieste,
I-34100 Italy\\
Istituto Nazionale di Fisica Nucleare, Sezione di Trieste, I-34127 Italy}
\maketitle
\begin{abstract}
\noindent We discuss the time dependent interferences between $K_L$ and $K_S$
in the decays in $3\pi$ and $\pi\pi\gamma$, to be studied at interferometry
machines such as the $\phi$-factory and LEAR. We emphasize the
possibilities and the advantages of using interferences, in comparison with
width measurements,
to obtain information both on $CP$ conserving and $CP$ violating amplitudes.
Comparison with present data and suggestions for future experiments are made.
\end{abstract}
\section{Introduction}
\label{sec:intro}
The origin of $CP$ violation is still an open problem in particle
physics. Particularly interesting in this regard is the field of kaon decays,
which is the only one where at least $CP$ violation in the mass-matrix has been
established through the measurements of $K_L\to\pi\pi$, $K_L\to\pi l\nu$ and
$K_L\to\pi\pi\gamma$ \cite{wolf1,paver1}. Further investigations are required
(both experimental and theoretical) to prove the existence of direct $CP$
violation, as required by the Standard Model. To this aim, and to elucidate
the mechanism of $CP$ violation, it is necessary to assess the potential
manifestations of $CP$ violation in kaon decay channels alternative to
$K^0\to\pi\pi$. This requires
a reliable theoretical approach to estimate the relevant hadronic matrix
elements of the nonleptonic $\Delta S=1$ weak Hamiltonian, and clearly
also calls for accurate experimental studies of kaon decays (both $CP$
violating and $CP$ conserving), to test the theoretical description. Of
course, an interesting aspect, besides the $CP$ violation problem, would be
the improvement of our understanding of meson dynamics and possibly a
clarification of some fundamental issues, such as {\it e.g.} the
origin of the $\Delta I=1/2$ rule.

In this paper we would like to discuss some selected examples,
emphasizing the role of experiments at machines such as $\phi$-factories
\cite{paver1,fukawa} and LEAR \cite{pavlo1}. The special feature of this kind
of machines is that they provide a well-defined initial $K_LK_S$ quantum
state, namely $\phi$-factories produce a $p$-wave $K^0{\overline{K^0}}$ state,
while tagged $K^0$ or ${\overline{K^0}}$ states are produced at LEAR.
Therefore, time-dependent interferences in the vacuum between $K_S$ and $K_L$
decaying to a final state $f$ can be accurately studied. For example, the
characteristic interference factor to be studied at LEAR
\cite{pavlo1,tanner,amelino} is of the form (assuming $CPT$):
\begin{equation}
2 e^{-\gamma t}\left[Re(\langle f\vert K_L\rangle^{\ast}
\langle f\vert K_S\rangle)\cos\Delta m t
-Im(\langle f\vert K_L\rangle^{\ast}
\langle f\vert K_S\rangle)\sin\Delta m t\right],\label{interf}\end{equation}
where $t$ is the proper time, $\gamma=(\gamma_L+\gamma_S)/2$ , and
$\Delta m = m_L-m_S$.
By fitting the time-dependence of Eq.(\ref{interf}) to the experimental data,
it is possible to determine independently both
${Re(\langle f\vert K_L\rangle^*\langle f\vert K_S\rangle)}$ and
${Im(\langle f\vert K_L\rangle^*\langle f\vert K_S\rangle)}$.

At the $\phi$-factory, where the initial state is produced {\it via} the decay
$\phi\to K^0{\overline{K^0}}$ with ${C=-1}$, one can measure
time-correlations between $K^0$ decay to a final state $f_1$ (used as a
tagging channel) at time $t_1$ and ${\overline{K^0}}$ decay to a state $f_2$
at time $t_2$. These correlations are expressed analogously to
Eq.(\ref{interf}), and in terms of the same physical quantities
\cite{fukawa,dunietz,peccei,dambrosio1}.

In what follows, we will discuss some interesting features of interferences in
neutral kaon decays to $3\pi$ and $\pi \pi \gamma$ final states, and
point out their relevance and physical implications. Interferences in these
channels should be measured with good statistics and precision at
Da$\phi$ne (the Frascati $\phi$-factory) or at LEAR. We will also emphasize
the complementary role of this kind of experimental analysis with respect to
width measurements.

Specifically, in Sec.\ \ref{sec:k3pi} we suggest that the recent LEAR
data should already enable us to put limits on the $3\pi$ final state
phases, which bring important information on the chiral structure of
meson-meson strong interactions and, even more important, determine the size
of direct $CP$ violation asymmetries in $K\rightarrow 3\pi$, so that their
experimental determination represents an essential piece of information. In
Sec.\ \ref{sec:kpipigamma}
we study $K_{L,S}\rightarrow \pi \pi \gamma$, and complement the
analysis of \cite{donoghue} by extending it to the case of LEAR and
including the electric ``direct emission'' $CP$ conserving amplitude, not
considered in that paper. In particular, we point out the relevance of
interference at both LEAR and Da$\phi$ne in determining such amplitude.
Finally, Sec.\ \ref{sec:concl} contains some concluding remarks.
\section{Notations and interference in $K\rightarrow 3\pi$}
\label{sec:k3pi}
We start by discussing the potential of LEAR, concerning the possibility
of measuring the $CP$ conserving $K_S\to 3\pi$ amplitude as well as the
final state ($3\pi$) phases. With the convention $\vert{\overline{K^0}}\rangle=
CP\vert K^0\rangle$, so that the eigenstates with definite $CP=\pm 1$ are
${\vert K_{1,2}\rangle=(\vert K^0\rangle\pm\vert{\overline{K^0}}\rangle)/
\sqrt 2}$,
the mass eigenstates are, assuming $CPT$ invariance as we shall do throughout
this paper:
\begin{equation}
\vert K_{S,L}\rangle=p\vert K^0\rangle\pm q\vert{\overline{K^0}}\rangle\equiv
\frac{\vert K_{1,2}\rangle+{\tilde\varepsilon}
\vert K_{2,1}\rangle}{\sqrt{1+\vert\tilde\varepsilon\vert^2}},
\label{masseigen}\end{equation}
where we adopt the same notations as in \cite{wolf1}. In particular:
\begin{equation}
{\tilde\varepsilon}=\varepsilon-i\frac{Im A_0}{Re A_0}\hskip 2pt
,\end{equation}
with $A_{0,2}$ the amplitudes for $K\to\pi\pi$ with $I=0,\hskip 2pt 2$.
Consequently, the proper time evolution of initial $K^0$ or
${\overline{K^0}}$ states is:
\begin{eqnarray}
\vert K^0(t)\rangle=\frac{\sqrt{1+\vert\tilde\varepsilon\vert^2}}
{\sqrt 2\left(1+
{\tilde\varepsilon}\right)}\left[\vert K_S\rangle
\exp{\left(\frac{-\Gamma_St}{2}-im_St\right)}+
\vert K_L\rangle
\exp{\left(\frac{-\Gamma_Lt}{2}-im_Lt\right)}\right],\nonumber\\
\vert{\overline{K^0}}(t)\rangle=\frac{\sqrt{1+\vert\tilde\varepsilon\vert^2}}
{\sqrt 2\left(1-
{\tilde\varepsilon}\right)}\left[\vert K_S\rangle
\exp{\left(\frac{-\Gamma_St}{2}-im_St\right)}-
\vert K_L\rangle
\exp{\left(\frac{-\Gamma_Lt}{2}-im_Lt\right)}\right].
\label{kevolution}\end{eqnarray}
At first order in $\varepsilon$ the amplitude squared for decay to a final
state $f$, as a function of time, is given in general by:
\begin{eqnarray}
\vert A\left(K^0({\overline{K^0}})\to f\right)\vert^2\simeq\frac{1}{2}
\left(1\mp 2Re\hskip 2pt\varepsilon\right)\big\{\exp{(-\Gamma_St)}
\vert A_S\vert^2+\exp{(-\Gamma_Lt)}\vert A_L\vert^2\nonumber \\
\pm 2\exp{(-\gamma t)}\left[Re\hskip 2pt\left(A_LA_S^*\right)
\cos{\Delta mt}+ Im\hskip 2pt\left(A_LA_S^*\right)\sin{\Delta mt}\right]
\big\},\label{a2evolution}\end{eqnarray}
where $\Delta m$ and $\gamma$ have already been defined in connection to
Eq.(\ref{interf}), and we use the notation
${A_{S,L}\equiv A(K_{S,L}\to f)}$.\par
Since we wish here to specialize Eq.(\ref{a2evolution}) to $K\to 3\pi$, we
adopt the familiar expansion of the amplitude for this process \cite{devlin}:
\begin{equation}
A(K_L\rightarrow\pi^+\pi^-\pi^0)=\left(\alpha_1+\alpha_3\right)
\exp{(i\delta_{1S})}
-\left(\beta_1+\beta_3\right)\exp{(i\delta_{1M})}Y
\label{kl3pi}\end{equation}
\begin{equation}
A(K_S\rightarrow\pi^+\pi^-\pi^0)=\frac{2}{\sqrt{3}}\gamma_3 X
\exp{(i\delta_2)}.\label{ks3pi}\end{equation}
In Eqs.(\ref{kl3pi}) and (\ref{ks3pi}), the subscripts 1 and 3 indicate
$\Delta I=1/2$ and $\Delta I=3/2$ transitions, respectively. Furthermore,
$X=(s_2-s_1)/m_{\pi}^2$ and $Y=(s_3-s_0)/m_{\pi}^2$ are the Dalitz plot
variables, with $s_i=(p_K-p_i)^2$, $s_0=(s_1+s_2+s_3)/3$ and $p_i$ the pions
momenta labelled in such a way that $i=3$ corresponds to the ``odd'' charge
pion (the $\pi^0$ in our case). The phases $\delta_{1S}$, $\delta_{1M}$
and $\delta_2$ originate from final-state strong interactions. The
experimental values of $\alpha_i$, $\beta_i$ and $\gamma_3$ have been obtained
from a fit to the experimental data on $K\to 3\pi$ differential widths
\cite{devlin,kambor}. The typical results, which can be used {\it e.g.} to
assess expected numbers of events, are
$\alpha_1+\alpha_3\simeq 8.5\times 10^{-7},
\ \beta_1+\beta_3\simeq -2.8\times 10^{-7},\ \gamma_3\simeq
2.3\times 10^{-8}$.

The strong phases are usually neglected in the fit and therefore are not
experimentally known yet. We recall that direct $CP$ violating asymmetries in
$K\to 3\pi$ are dominated by the interferences between the $\Delta I=1/2$
amplitudes $\alpha_1$ and $\beta_1$ and between $\alpha_1$ and the
$\Delta I=3/2$ amplitude $\gamma_3$. Consequently, these asymmetries
are proportional to $\delta_{1S}-\delta_{1M}$ and $\delta_{1S}-\delta_2$.
Therefore, the experimental determination of these phases is crucial to make
estimates for direct $CP$ violation in this decay
channel \cite{wolf1,isidori}. Another interesting aspect of this
determination of the low energy ($3\pi$) phases
is that it would usefully complement the measurement of $\pi$-$\pi$ phase
shifts near threshold in {\it e.g.} $K_{l4}$ decays or $\pi$-$\pi$
scattering, and thus would allow a stringent test of the current theoretical
approach to meson dynamics, based on effective chiral Lagrangians. As
emphasized in Sec.\ \ref{sec:intro} and in Ref.\cite{dambrosio1}, in this
regard
the unique advantage of interference is that it depends linearly on the
($3\pi$) strong phases, which are expected to be small. For example, at the
centre of the Dalitz plot, the theoretical expectations from both chiral loops
\cite{isidori} and a nonrelativistic approach \cite{zeldovich} are:
\begin{equation}
\delta_{1S}-\delta_{1M}\simeq\frac{\delta_{1S}-\delta_2}{2}\simeq 0.07
\end{equation}
Conversely,  width measurements only give $\cos\delta\simeq 1-\delta^2/2$ and
thus are less sensitive to small $\delta$.

An analogous linear dependence {\it vs} a quadratic
one occurs also for ${A(K_S\to\pi^+\pi^-\pi^0)}$: one can see
from (\ref{ks3pi}) that the width of this process is suppressed both by the
$\Delta I=1/2$ rule and by the angular momentum barrier. One can remark
that linear dependence on the ($3\pi$) phases $\delta$ and on the amplitude
$\gamma_3$ could also obtain in a regenerator experiment \cite{ford}, but
compared to the possibility of observing oscillations in vacuum at LEAR or
Da$\phi$ne this would be affected by a significant uncertainty due to the
regeneration parameters.

The experimental analysis of the interference should proceed by substituting
Eqs.(\ref{kl3pi}) and (\ref{ks3pi}) in the time evolution
Eq.(\ref{a2evolution}), making suitable kinematical cuts over the Dalitz
plot, and fitting to the experimental time dependence
\cite{amelino,dambrosio1}. In the specific case, to determine the
$CP$ conserving amplitude of $K_S\to 3\pi$, which according to (\ref{ks3pi})
is antisymmetric in $X$,
the following weighted integral over the Dalitz plot can be considered
\cite{amelino,dambrosio1}:
\begin{eqnarray}
A^{+-0}_X (t)=\frac{\int d\Phi\ sgn (X)
\left[\vert A(K^0\to\pi^+\pi^-\pi^0)\vert^2-
\vert A({\overline{K^0}}\to\pi^+\pi^-\pi^0)\vert^2\right]}
{\int d\Phi\left[\vert A(K^0\to\pi^+\pi^-\pi^0)\vert^2+
\vert A({\overline{K^0}}\to\pi^+\pi^-\pi^0)\vert^2\right]}\nonumber\\
=
\frac{a
\gamma_3(\alpha_1+\alpha_3)\exp{(-\gamma t)}
\left[\cos(\delta_{1S}-\delta_2)\cos{\Delta mt}+
\sin(\delta_{1S}-\delta_2)\sin{\Delta mt}\right]}
{g\gamma_3^2  \exp{(-\Gamma_S t)}+
[(\alpha_1+\alpha_3)^2+b(\beta_1+\beta_3)^2 ]
\exp{(-\Gamma_L t)}},
\label{asym3pi}\end{eqnarray}
where $d\Phi$ denotes the phase space element. In Eq.(9) $a,g$ and $b$ come
from phase space integrations, and their values are:
\begin{equation}
a=\frac{32 m_K Q }
{9\pi m_\pi^2}\hskip 2pt ;\quad
g=\frac{4m_K^2 Q^2}{9m_\pi^4}\hskip 2pt ;\quad
b=\frac{m_K^2 Q^2}{9m_\pi^4}\hskip 2pt ,\label{psf}
\end{equation}
with $Q$ the $Q$-value of the reaction, $Q^{+-0}=83.6\hskip 2pt MeV$.
The denominator in (\ref{asym3pi}) has been chosen just to conveniently
normalize the asymmetry, although different choices are quite possible. Also,
one can notice that in the denominator of (\ref{asym3pi}) only the second
term is numerically relevant.

Recently, LEAR has produced a preliminary, direct determination of $\gamma_3$
\cite{montanet} from a fit of (\ref{asym3pi}), neglecting the
$\sin\Delta m t$ part. Actually, we would suggest experimentalists to fit the
data with the full Eq.(\ref{asym3pi}), and derive (at least) an upper limit on
$\delta_{1S}-\delta_2$. Naively, by just imposing that the neglected term
should be less or equal to the quoted statistical error of 30\%
on the $K_S\to3\pi$
amplitude, one would expect an upper limit on this combination of phases of
the order of $30^\circ$. This is already at the level of discarding models
claiming enhanced $CP$ violating asymmetries from large ($3\pi$) strong phases
\cite{belkov}.

An analogous analysis can be performed to obtain a
bound on the independent combination $\delta_{1S}-\delta_{1M}$, by
considering a cut in $X\cdot Y$ \cite{dambrosio1}:
\begin{eqnarray}
A^{+-0}_{XY} (t)=\frac{\int d\Phi\ sgn (X\cdot Y)
\left[\vert A(K^0\to\pi^+\pi^-\pi^0)\vert^2-
\vert A({\overline K^0}\to\pi^+\pi^-\pi^0)\vert^2\right]}
{\int d\Phi\left[\vert A(K^0\to\pi^+\pi^-\pi^0)\vert^2+
\vert A({\overline K^0}\to\pi^+\pi^-\pi^0)\vert^2\right]}\nonumber\\
=\frac{ -h\gamma_3(\beta_1+\beta_3)\exp{(-\gamma t)}
\left[\cos(\delta_{1S}-\delta_{1M})\cos{\Delta mt}+
\sin(\delta_{1S}-\delta_{1M})\sin{\Delta mt}\right]}
{g\gamma_3^2\exp{(-\Gamma_S t)}+
[(\alpha_1+\alpha_3)^2+b (\beta_1+\beta_3)^2 ]
\exp{(-\Gamma_L t)}}
\label{asymm3pi}\end{eqnarray}
where $h$ is the another phase space integral
\begin{equation}
h=\frac{8 m_K^2 Q^2 }{9\pi m_\pi^4}\hskip 2pt .\label{psf2}
\end{equation}
\par In the future, the foreseen improvement in the experimental uncertainty
(a factor 4 in statistics) should open the way to precise measurements of
the $K\to3\pi$ phases. Concerning $CP$ violation in this decay channel,
the leading effect proportional to $Re\hskip 2pt\varepsilon$ in
Eq.(\ref{a2evolution}) is obtained without making any cut on the Dalitz plot,
because it is symmetric in $X$. Also, bounds on direct $CP$ violation can be
obtained in principle. Considering the expected extreme smallness of this
effect and the present capabilities, presumably this will need a really new
stage in experimental accuracy.\par
Turning to the $\phi$-factory, the initial quantum state is represented in
this case by the superposition of $K_L$ and $K_S$ states:
\begin{equation}
\vert i\rangle\equiv \vert K^0 {\overline{K^0}}(C=odd)\rangle=
\frac{\vert K_{L}(\hat{z}) K_{S}(-\hat{z})\rangle
-\vert K_{S}(\hat{z}) K_{L}(-\hat{z})\rangle}{2\sqrt{2} pq}\hskip 2pt ,
\label{phi}
\end{equation}
where $\hat z$ is the direction of kaons momenta in the c.m. system.
The subsequent $K_L$ and $K_S$ decays are correlated,
and their quantum interferences show up in relative
time distributions and time asymmetries.
Specifically, one considers the transition amplitude for the
initial state to decay into the final states $f_1$ at time $t_1$ and
$f_2$ at time $t_2$, respectively. Defining the ``intensity'' of
time-correlated events $I(\Delta t)$ as:
\begin{equation}
I(f_1,f_2;\Delta t)\equiv {1\over 2} \int\limits^{\infty}_{\vert\Delta t\vert}
dt\hskip 3pt
\vert\langle f_1(t_1,\hat{z}),f_2(t_2,-\hat{z})\vert T \vert i\rangle
\vert^2,\label{intensity}\end{equation}
where $t=t_1+t_2$ and $\Delta t=t_2-t_1$, we find:

\begin{eqnarray}
I(\Delta t<0)&=&
\Big\{\exp{(-\Gamma_S\vert\Delta t\vert)}
\vert A_S(f_1)\vert^2\vert A_L(f_2)\vert^2
+\exp{(-\Gamma_L\vert\Delta t\vert)}
\vert A_L(f_1)\vert^2 \vert A_S(f_2)\vert^2 \nonumber\\
&-&2\exp{(-\gamma\vert\Delta t\vert)}
\Bigl[Re\hskip 2pt\left(A_L(f_1)A_S(f_1)^* A_L(f_2)^*A_S(f_2)\right)
\cos{\Delta m\vert\Delta t\vert}\nonumber\\
&+&Im\hskip 2pt\left(A_L(f_1)A_S(f_1)^* A_L(f_2)^*A_S(f_2)\right)
\sin{\Delta m\vert\Delta t\vert}\Bigr]
\Big\}\frac{1}{16\gamma\vert p\vert^2\vert q\vert^2},\label{ievolution}
\end{eqnarray}
and

\begin{eqnarray}
I(\Delta t>0)&=&
\Big\{\exp{(-\Gamma_L\Delta t)}
\vert A_S(f_1)\vert^2\vert A_L(f_2)\vert^2
+\exp{(-\Gamma_S\Delta t)}
\vert A_L(f_1)\vert^2 \vert A_S(f_2)\vert^2 \nonumber\\
&-&2\exp{(-\gamma\Delta t)}
\Bigl[Re\hskip 2pt\left(A_L(f_1)A_S(f_1)^* A_L(f_2)^*A_S(f_2)\right)
\cos{\Delta m\Delta t}\nonumber\\
&-&Im\hskip 2pt\left(A_L(f_1)A_S(f_1)^* A_L(f_2)^*A_S(f_2)\right)
\sin{\Delta m\Delta t}\Bigr]
\Big\}\frac{1}{16\gamma\vert p\vert^2\vert q\vert^2}.\label{iievolution}
\end{eqnarray}
Here, we denote $A_{S,L}(f_i)\equiv A(K_{S,L}\to f_i)$ with $i=1,2$.
The theoretical analysis of the $CP$ conserving amplitude of $K_S\to 3\pi$,
based on the time correlations (\ref{ievolution}) and (\ref{iievolution}),
was presented in \cite{dambrosio1} with the choice $f_1=\pi l\nu$ as the
tagging process and $f_2=\pi^+\pi^-\pi^0$. $CP$ violation in the mass matrix
can be studied similar to the case of Eq.(\ref{a2evolution}), either with the
same $f_1$ and $f_2$ as in \cite{dambrosio1} or with
$f_1=f_2=3\pi$ \cite{donoghue}, and according to the previous discussion
this can be regarded as an alternative to the measurement of
$\Gamma(K_S\to3\pi^0)$. Direct $CP$ violation can also be studied, although
the effect is predicted to be so small in the Standard Model that presumably
at best an upper limit could be achieved.
\section{Interference in $K\to\pi\pi\gamma$}
\label{sec:kpipigamma}
The amplitudes for $K\rightarrow \pi\pi\gamma$ decays are generally defined as
the superposition of two amplitudes: internal bremsstrahlung $(A_{IB})$ and
direct emission $(A_{DE})$ \cite{report}. $A_{IB}$ accounts for
bremsstrahlung from external charged particles and is predicted
simply by QED in terms of the $K\to\pi\pi$ amplitude. $A_{DE}$ is obtained by
subtraction of $A_{IB}$ from the total amplitude, and accounts for the
possibility of direct $K\rightarrow\pi\pi\gamma$ couplings. By definition
this amplitude is a test for mesonic interaction models, and in particular for
the current description based on effective Lagrangians and Chiral
Perturbation Theory (ChPT) \cite{gasser}.

For the processes
$K_{S,L}(p_K)\rightarrow \pi^+(p_+) \pi^-(p_-) \gamma(q,\epsilon)$,
we write
\begin{equation} A^{S,L}=A^{S,L}_{IB}+A^{S,L}_{DE}\end{equation}
where
\begin{equation}
A^{S,L}_{IB}=eB\hskip 2pt A(K_{S,L}\rightarrow\pi^+\pi^-)\label{ib}
\end{equation}
and
\begin{equation}
A^{S,L}_{DE}=e{\bar{B}}\hskip 2pt h_E^{S,L}(E_{\gamma}^*,\cos\theta)
A(K_{S}\rightarrow\pi^+\pi^-)+
eB_M\hskip 2pt h_{M}^{S,L}(E_{\gamma}^*,\cos\theta).\label{fde}\end{equation}
Here, $E_\gamma^*$ is the photon energy in the $K_{S,L}$ rest frame and
$\theta$ is the angle between the photon and the $\pi^+$ in the dipion frame.
Furthermore:
\begin{eqnarray}
B&=&\frac{\epsilon \cdot p_+}{ q \cdot p_+}  -
\frac{\epsilon \cdot p_-}{ q \cdot p_-},\nonumber \\
{\bar{B}}& = &\epsilon \cdot p_+ ~q \cdot p_- ~-~
\epsilon \cdot p_-~ q \cdot p_+ \nonumber\\
B_M&=&\varepsilon_{\alpha\beta\gamma\delta} p_+^\alpha p_-^\beta q^\gamma
\epsilon^\delta.\label{b}\end{eqnarray}
These are the only possible gauge and Lorentz invariant structures up to
third order in momenta. While $B$ corresponds to the $IB$ amplitude,
${\bar{B}}$ and $B_M$ correspond to electric and magnetic transitions,
respectively. If the photon polarization is not measured there is no
interference among electric and magnetic transitions, so that the
differential width is
\begin{equation}
d\Gamma=d\Gamma_{IB}+d\Gamma_{int}+d\Gamma_{M}+d\Gamma_{|{\overline{B}}|^2},
\label{digamma}\end{equation}
where $int$ represents the interference between the $IB$ and the $DE$
electric components. On the r.h.s of Eq.(\ref{fde}) we have explicitly
factored $A(K_S\to\pi^+\pi^-)$ for later convenience, in order that all
quantities of interest have a common factor
${\vert A(K_S\to\pi^+\pi^-)\vert^2}$.
Of course, in using our formulae involving $h_E^{S,L}$ and comparing with
direct emission amplitudes defined in the literature, this normalization has
to be taken into account.

Concerning $CP$ violation in this radiative nonleptonic process, and direct
$CP$ violation in particular, we recall
that in the limit of $CP$ conservation, and to the lowest contributing
multipoles, for the $K_S$ decay $h_M^S=0$ and the direct emission amplitude
is determined by $h_{E1}^S$ (electric dipole moment E1), whereas $K_L$
decay proceeds only through $h_{M1}^L$ (magnetic dipole moment M1)
\cite{report}. However, reflecting the bremsstrahlung enhancement for
$E_{\gamma}^*\to 0$, the $CP$ violating $A_{IB}^L$  is experimentally found
to compete with the $CP$ conserving $DE$ amplitude $h_M^L$
\cite{barker,ramberg}.
According to (\ref{ib}) $A_{IB}^L$ is proportional to $A(K_L\to\pi^+\pi^-)$,
where the direct $CP$ violation parameter ${\epsilon^\prime_{\pi\pi}}$ is
very small, being suppressed by the $\Delta I=1/2$ rule. Therefore, this
direct $CP$ violation effect can hardly show up in the rate (\ref{digamma})
for $K_L\to\pi^+\pi^-\gamma$. Conversely, direct $CP$ violation in  $h_E^L$ is
not related to $K_L\to\pi\pi$, and therefore in principle might give a larger
effect through the $int$ term in (\ref{digamma}). However, unfortunately the
$IB$ enhancement cannot be beaten, and in fact such interference is expected
to be strongly suppressed relative to the $IB$ term. This suppression factor
can be qualitatively guessed {\it e.g.} by considering the total
interferencial width between  $A^{S}_{IB}$ and
${e{\bar {B}}\hskip 2pt h_E^{S}(E_{\gamma}^*,\cos\theta)
A(K_{S}\to\pi^+\pi^-)}$. For simplicity of notations we denote it as follows:
\begin{equation}
\langle Re h_E^S \rangle_{int}
\equiv e^2\vert A(K_{S}\rightarrow\pi^+\pi^-)\vert^2
\int d\Phi\hskip 2pt (\sum_{Pol} B {\bar{B}})\hskip 2pt Re(h_E^S),
\label{Reinterf}\end{equation}
where $\Phi$ is the phase space, which must be cut for experimentally
undetected photons.
As shown in Ref.\cite{dambrosio2}, at $O(p^4)$ in ChPT $h_E^S$
is the sum of a predicted  loop amplitude and an unknown counterterm.
Taking $E_{\gamma}^* >20\ MeV$  and
varying the counterterm in a reasonable range one obtains a negative
interference, of the order of \cite{dambrosio2}:
\begin{equation}
\frac{\Gamma(K_{S}\rightarrow\pi^+\pi^-\gamma)_{int}}
{\Gamma(K_{S}\rightarrow\pi^+\pi^-\gamma)_{IB}}\simeq -10^{-4}\div -10^{-3},
\label{factor}\end{equation}
or, using $Br(K_S\to\pi^+\pi^-\gamma, E_\gamma^*> 20 MeV)_{IB}\simeq
4.8 \times 10^{-3}$:
\begin{equation}Br(K_{S}\rightarrow\pi^+\pi^-\gamma)_{int}\simeq
-10^{-6}\div -10^{-5}.\label{suppression}\end{equation}
This result strongly suppresses the expected sensitivity of the
${K_L\to\pi\pi\gamma}$ width measurement to the $CP$ violating amplitude
$h^L_E$, so that time-dependent interference analysis could represent a
viable alternative in this case.

In general, besides the $CP$ violation problem, a stringent test of the
theoretical framework for the relevant hadronic matrix elements of $H_W$
could conveniently be performed in such interference experiments. Indeed,
one interesting point concerning $A_{DE}^L$ is that higher multipoles are in
general not suppressed by $CP$ violation, and might interfere significantly
with $A_{IB}^{S,L}$. For example, being $h^{S,L}_{E2}$ odd under
$\pi^+\leftrightarrow\pi^-$ interchange ($\theta\to\theta+\pi$), by a
suitable kinematical cut at the $\phi$-factory or at LEAR one could project
out the interference between this amplitude and $A_{IB}$.

Regarding the current theoretical situation, both the lowest multipole
amplitudes and the higher ones can be predicted in ChPT to order
$p^4$ \cite{enp}. Specifically, the $CP$ violating $O(p^4)$ amplitude
$h_E^L$ is expressed, in that framework, in terms of meson loops
($h^L_{E,{ l}}$), which are related to $CP$ violation in $K^0\to\pi\pi$,
and of matrix elements of local counterterm operators ($h^L_{E,ct}$) not
suppressed by the $\Delta I=1/2$ rule. For the case of
$K_S\to\pi^+\pi^-\gamma$, the $CP$ conserving $O(p^4)$ direct emission
amplitude $h^S_E$ is expressed
analogously, in terms of loops ($h^S_{E,{l}}$) and counterterms
($h^S_{E,ct}$) \cite{dambrosio2}. The same counterterms contribute
to ${K_1\to\pi^+\pi^-\gamma}$ as well as to ${K_2\to\pi^+\pi^-\gamma}$,
with coefficients that are,
respectively, real or imaginary in the chosen phase convention for $K^0$ and
${\overline{K^0}}$. Similar to the case of ${K\to 3\pi}$ \cite{isidori2},
the direct $CP$ violating component of the
loop amplitudes $h^L_{E,{l}}$ can be obtained by multiplying the
$h^S_{E,{l}}$ by a factor containing the combination of CKM angles
appropriate to ${\varepsilon^\prime_{\pi\pi}/\varepsilon}$. Conversely, such a
simple relation does not hold between the contributions of $CP$ conserving and
$CP$ violating counterterms, because the actual values of the
counterterm coefficients cannot be related, in general, to the process
$K_L\rightarrow \pi \pi$.

For the $CP$ violating $DE$ amplitude we will use in the sequel the following
parametrization of $h^L_{E1}$:
\begin{eqnarray}
h_{E1}^L&\equiv&\frac{A( K_L\rightarrow \pi^+ \pi^-\gamma)_{DE,E1}}
{eA( K_S\rightarrow \pi^+ \pi^-){\bar{B}}}\nonumber\\
&\simeq&\left(\varepsilon-i\frac{ImA_0}{ReA_0}\right)h^S_{E1}+
\frac{A( K_2\rightarrow \pi^+ \pi^-\gamma)_{E1}}
{eA( K_S\rightarrow \pi^+ \pi^-){\bar{B}}}\simeq\varepsilon h^S_{E1}+
\varepsilon_{\pi\pi\gamma}^\prime h^S_{E1,ct},
\label{hl}\end{eqnarray}
where for phenomenological purposes we have neglected
$\varepsilon^{\prime}_{\pi\pi}$ with respect to $\varepsilon$. In (\ref{hl})
$\varepsilon_{\pi\pi\gamma}^\prime h^S_{E1,ct} $ is the genuine,
direct $CP$ violation
term in $K^0\to\pi\pi\gamma$. We have normalized it in such a way that,
if this direct $CP$ violation is generated by the $O(p^4)$ local
counterterms as we expect, then $\varepsilon_{\pi\pi\gamma}^\prime$ does
not depend on energy. Of course, this is a theoretical bias.

To discuss the opportunities of measurements at LEAR, we consider the following
asymmetry as a function of proper time:
\begin{eqnarray}
A_{\pi^+\pi^-\gamma}&\equiv&\displaystyle{
\frac{\int d\Phi\left[\vert A(K^0\rightarrow \pi^+\pi^-\gamma)\vert^2
-\vert A({\overline{K^0}}\rightarrow \pi^+\pi^-\gamma)\vert^2\right]}
{\int d\Phi\left[\vert A(K^0\rightarrow \pi^+\pi^-\gamma)\vert^2+
\vert A({\overline{K^0}}\rightarrow \pi^+\pi^-\gamma)\vert^2\right]}
\cong -2Re\varepsilon}\nonumber\\
&+&
{\displaystyle\frac{\left(2\exp{(-\gamma t)}/\Gamma_S\right)}
{\exp{(-\Gamma_S t)}
Br(K_{S}\rightarrow\pi^+\pi^-\gamma)+\frac{\Gamma_L}{\Gamma_S}
\exp{(-\Gamma_L t)}Br(K_{L}\rightarrow\pi^+\pi^-\gamma)
}}\nonumber \\
&\times &\Big\{\Gamma_S Br(K_{S}\rightarrow\pi^+\pi^-\gamma)_{IB}
[Re\eta_{+-}\cos{\Delta mt}+Im\eta_{+-}\sin{\Delta mt}]\nonumber \\
&+&
\langle Re({h_{E1}^{S*}}\eta_{+-}+h_{E1}^L)\cos{\Delta mt}+
Im({h_{E1}^{S*}}\eta_{+-}+h_{E1}^L)\sin{\Delta mt}\rangle_{int}
\Big\},
\label{APP}\end{eqnarray}
where the notation $\langle\cdots \rangle_{int}$ has the same
meaning as in (\ref{Reinterf}), and we have introduced the familiar ratio
${\eta_{+-}=A(K_L\to\pi^+\pi^-)/A(K_S\to\pi^+\pi^-)}$. We have limited to the
lowest significant $DE$ multipoles, and have neglected the small pure $DE$
emission contributions. Also, we have neglected in the denominator the
interference term, suppressed by a $CP$ violation factor and always
numerically smaller than the other ones.
In (\ref{APP}), we have the mass-matrix  $CP$ violation term
($-2Re\hskip 2pt \varepsilon$), and three different time-oscillating terms.

The first time-dependent term is the one proportional to:
\begin{equation}
Br(K_{S}\rightarrow\pi^+\pi^-\gamma)_{IB}[Re\eta_{+-}\cos{\Delta mt}+
Im\eta_{+-}\sin{\Delta mt}],\label{IBt}\end{equation}
which measures the $CP$ violation in ${K\rightarrow\pi\pi\gamma}$ related to
${K_L\to\pi\pi}$. With the value of $Br(K_S\to\pi^+\pi^-\gamma)_{IB}$
leading to Eq.(\ref{suppression}) and
$Re\hskip 2pt \eta_{+-}\simeq Im\hskip 2pt\eta_{+-}\sim 10^{-3}$,
integrating the intensity of events
over time (essentially a few $\tau_S$) we find that with
$10^8$ initial $K^0$ or ${\overline{K^0}}$ one can expect about 100 events
related to this term. We emphasize that, although this source of $CP$
violation is known as being related by QED to the one in $K_L\to\pi\pi$,
such a measurement is still
interesting because it allows to establish $CP$ violation in a different decay
channel. Actually, this analysis might be competitive
with the one  at Fermilab \cite{barker}, measuring $CP$ violation in
\begin{equation}
\vert\eta_{+-\gamma}\vert
=\Big\vert\frac{A( K_L\rightarrow \pi^+ \pi^-\gamma)_{IB+E1}}
{A( K_S\rightarrow \pi^+ \pi^-\gamma)_{IB+E1}}\Big\vert=
(2.15\pm0.26\pm0.20)\times 10^{-3},\label{etagamma}
\end{equation}

\begin{equation}
\phi_{+-\gamma}=\arg\Big\{\frac{A( K_L\rightarrow \pi^+ \pi^-\gamma)_{IB+E1}}
{A( K_S\rightarrow \pi^+ \pi^-\gamma)_{IB+E1}}\Big\}=(72\pm 23\pm 17)^\circ
,\label{phigamma}
\end{equation}
assuming  constant $h_{E1}^{S,L}$. From the experimental findings
${\arg{\varepsilon}=(43.67\pm 0.14)^\circ}$ and
${\vert\varepsilon^{\prime}_{\pi\pi}/\varepsilon\vert\le 2.3 \times 10^{-3}}$
\cite{wolf1}, combining with the theoretical expectation
${\arg\varepsilon^{\prime}_{\pi\pi}=(43\pm 6)^\circ}$ \cite{ochs}, it seems
reasonable to assume the approximation ${Re\hskip 2pt\eta_{+-}\cong
Im\hskip 2pt\eta_{+-}}$, so that the term (\ref{IBt}) has the characteristic
time dependence:
\begin{equation}
Re\hskip 2pt\eta_{+-}\hskip 2pt (\cos{\Delta m t}+\sin{\Delta m t})
\equiv\sqrt{2}\hskip 2pt Re\hskip 2pt\eta_{+-}\hskip 2pt\sin{(\Delta
mt+\pi/4)}.
\label{IBT}
\end{equation}
Clearly, for a more accurate estimate, one can easily include in (\ref{IBT})
the small deviation of ${\arg\varepsilon}$ from ${45^\circ}$.

More interesting is the second part of the interference term in
Eq.(\ref{APP}), which according to (\ref{hl}) can be split in two parts.
The first one is
\begin{equation}
2Re\hskip 2pt\varepsilon\langle Re\hskip 2pt h_{E1}^S
(\cos{\Delta m t}+\sin{\Delta m t})\rangle
_{int},\label{fdet}\end{equation}
in the approximation ${\eta_{+-}\simeq\varepsilon}$ and
${Im\hskip 2pt\varepsilon\simeq Re\hskip 2pt\varepsilon}$.
Eq.(\ref{fdet}) has the same time dependence as the IB term (\ref{IBT}), and
therefore could be distinguished only by looking at the $E_{\gamma}^*$ and
$\cos\theta$  dependence, similar to the rate measurement of
Eq.(\ref{digamma}). Thus, by observing this time correlation one can measure
$\langle Re\hskip 2pt h_{E1}^S\rangle$, i.e. the interference
between $A_{IB}^L$ and
$A_{E1}^S$. This is the same interference term obtained in the width
measurement. With only $10^8$ initial ${K^0({\overline{K^0}})}$ the
corresponding rate of events might be too low, but anyway one could
significantly improve existing bounds on the $DE$ amplitude
for $K_S$ decay.\par
Finally, there is the term which can measure the direct CP violation parameter
$\varepsilon_{\pi\pi\gamma}^\prime$:
\begin{equation}
\langle Re(\varepsilon_{\pi\pi\gamma}^\prime h_{E1,ct}^S)
\cos{\Delta m t}+
Im(\varepsilon_{\pi\pi\gamma}^\prime h_{E1,ct}^S)\sin{\Delta m t}\rangle_{int}.
\label{cpt}\end{equation}
The important point is that this term has a different time behavior, compared
to (\ref{IBT}) and (\ref{fdet}), and consequently should be disentangled
from the others by accurate time dependence studies. Taking into account the
imaginary character of the counterterm coefficients, one can notice that
$\varepsilon_{\pi\pi\gamma}^\prime h_{E1,ct}^S$ should have
a real part, due to our definition (\ref{fde}) and final state strong
interactions which make ${A(K_S\to\pi^+\pi^-)}$ complex.
Although some theoretical models seem to indicate a suppression of
$\varepsilon_{\pi\pi\gamma}^\prime$ \cite{cheng}, we nevertheless believe
that a value of the order of $10^{-4}\sim 10^{-5}$ is not completely
unreasonable.
Then, multiplying Eq.(\ref{cpt}) by the interference factor $\sim 10^{-5}$
of Eq.(\ref{suppression}), we see that it should be possible to put
interesting bounds on this direct $CP$ violation parameter with
${10^{8}}$-${10^{9}}$ initial ${K^0({\overline{K^0}})}$.
Probably, this can be more easily done by studying the interference in the
kinematical region where the $IB$ contribution is less important, which
is the region of maximum photon energy.

Interferometry experiments might provide also a unique method to detect
higher multipole transition amplitudes {\it via} the interference with the
$IB$ amplitude. For example, by considering the following cut in the Dalitz
plot:
\begin{equation}
\int d\Phi^{\theta}\equiv
\int d\Phi \ sgn (\sin{\theta}),\label{theta}\end{equation}
one can define an asymmetry to extract $h^L_{E2}$:
\begin{eqnarray}
A_{\pi^+\pi^-\gamma}^{\theta}&\equiv&
\frac{\int d\Phi^{\theta}\left[\vert A(K^0\rightarrow \pi^+\pi^-\gamma)\vert^2
-\vert A({\overline{K^0}}\rightarrow \pi^+\pi^-\gamma)\vert^2\right]}
{\int d\Phi\left[\vert A(K^0\rightarrow \pi^+\pi^-\gamma)\vert^2+
\vert A({\overline{K^0}}\rightarrow \pi^+\pi^-\gamma)\vert^2\right]}\cong
\frac{2\exp{(-\gamma t)}}{\Gamma_S}\nonumber \\
&\times&\frac{\langle Re\hskip 2pt h_{E2}^L\cos{\Delta mt}+
Im\hskip 2pt h_{E2}^L\sin{\Delta mt}
\rangle_{int}^\theta }{\exp{(-\Gamma_S t)}
Br(K_S\to\pi^+\pi^-\gamma)+\frac{\Gamma_L}{\Gamma_S}
\exp{(-\Gamma_L t)}Br(K_L\to\pi^+\pi^-\gamma)}.\nonumber \\
&&
\label{APP-}\end{eqnarray}
Here, with an obvious extension of the notation in Eq.(\ref{Reinterf}), we
have introduced the interference:
\begin{equation}
\langle Re\hskip 2pt h_E^L \rangle_{int}^\theta
\equiv e^2\vert A(K_{S}\rightarrow\pi^+\pi^-)\vert^2
\int d\Phi\hskip 2pt (\sum_{Pol} B {\bar{B}})\hskip 2pt sgn (\sin{\theta})
\hskip 2pt Re\hskip 2pt h_E^L\hskip 2pt .\label{Reinterf-}\end{equation}
To derive Eq.(\ref{APP-}), we have used the well-known fact that
${\vert A_{IB}\vert^2}$ is even in $\sin\theta$ and ${h_E^S}$ is independent of
$\sin\theta$ to ${O(p^4)}$ in ChPT, and have neglected terms proportional to
$CP$ violation.

The $O(p^4)$ one-loop $CP$ conserving amplitude
${h_{E2}^L A(K_S\rightarrow\pi^+\pi^-)}$ has been computed in ChPT
 \cite{enp}. In this theoretical framework, this product does not have
absorptive part, i.e. it should be
purely real. By looking at the values of this amplitude
over the Dalitz plot, the authors of Ref.\cite{enp} suggest
\begin{equation}
\vert \frac{e h^L_{E2} A(K_S\rightarrow \pi^+\pi^-)\bar{B}}
{A(K_{L}\rightarrow\pi^+\pi^-\gamma)_{IB}}\vert\le 10^{-2}
\hskip 2pt .\label{hl2}\end{equation}
As mentioned above, due to final state interactions
${A(K_S\rightarrow \pi^+\pi^-)}$ has both real and imaginary parts.
Consequently, both terms in (\ref{APP-})
are present and possibly might be separately measured. Assuming
optimistically the
upper bound (\ref{hl2}) to be saturated all over the Dalitz plot,
with $10^8$-$10^9$ initial kaons few events should be available
for this kind of analysis.

At the $\phi$-factory, it is possible to study $K\rightarrow \pi \pi \gamma$
decays through interferences by choosing in Eqs.(\ref{ievolution}) and
(\ref{iievolution}) ${f_1=\pi^+\pi^-}$, ${\pi^0\pi^0}$ or
${\pi^{\pm} l^{\mp} \bar{\nu} (\nu)}$ as tagging channels, and
${f_2=\pi^+ \pi^-\gamma}$.

For the case ${f_1=\pi^+\pi^-}$ one obtains:
\begin{eqnarray}
I(\Delta t < 0 )_{\scriptstyle \pi^+ \pi^-}
&=&\frac{\Gamma(K_S\rightarrow \pi^+\pi^-)
\Gamma(K_S\rightarrow \pi^+\pi^-\gamma)_{IB}}
{16\gamma\vert p\vert^2\vert q\vert^2}\cr
&\times&\Big\{\exp{(-\Gamma_S\vert\Delta t\vert)}\hskip 2pt R_L
+\exp{(-\Gamma_L\vert\Delta t\vert)}
\vert\eta_{+ -}\vert^2\hskip 2pt R_S\cr
&-& 2\exp{(-\gamma\vert\Delta t\vert)}
\Bigl[\Bigl(\vert\eta_{+ -}\vert^2+\frac
{Re(\eta_{+-}\langle h^{L*}_{E1}+\eta^*_{+-} h_{E1}^S\rangle _{int})}
{\Gamma(K_S\to\pi^+\pi^-\gamma)_{IB}}
\Bigr)\cos{\Delta m\vert\Delta t\vert}\cr
&+& Im\Bigl(\eta_{+ -}
\frac{\langle h^{L*}_{E1}+\eta^*_{+-} h_{E1}^S\rangle_{int}}
{\Gamma(K_S\to\pi^+\pi^-\gamma)_{IB}}\Bigr)
\sin{\Delta m\vert\Delta t\vert}\Bigr]\Big\},\label{ievolutionkppg}
\end{eqnarray}

\begin{eqnarray}
I(\Delta t > 0 )_{\scriptstyle \pi^+ \pi^-}
&=&\frac{\Gamma(K_S\rightarrow \pi^+\pi^-)
\Gamma(K_S\rightarrow \pi^+\pi^-\gamma)_{IB}}
{16\gamma\vert p\vert^2\vert q\vert^2}\cr
&\times&\Big\{\exp{(-\Gamma_L\Delta t)}\hskip 2pt R_L
+\exp{(-\Gamma_S\Delta t)}
\vert\eta_{+ -}\vert^2\hskip 2pt R_S\cr
&-& 2\exp{(-\gamma\Delta t)}
\Bigl[\Bigl(\vert\eta_{+ -}\vert^2+\frac
{Re(\eta_{+-}\langle h^{L*}_{E1}+\eta^*_{+-} h_{E1}^S\rangle _{int})}
{\Gamma(K_S\to\pi^+\pi^-\gamma)_{IB}}
\Bigr)\cos{\Delta m \Delta t}\cr
&-& Im\Bigl(\eta_{+ -}
\frac{\langle h^{L*}_{E1}+\eta^*_{+-} h_{E1}^S\rangle_{int}}
{\Gamma(K_S\to\pi^+\pi^-\gamma)_{IB}}\Bigr)
\sin{\Delta m \Delta t} \Bigr]\Big\},\label{iievolutionkppg}
\end{eqnarray}
and similar formulae hold for $f_1=\pi^0\pi^0$. For convenience we have
introduced the ratios:
\begin{equation}
R_L=\frac{\Gamma(K_L\rightarrow \pi^+\pi^-\gamma)}
{\Gamma(K_S\rightarrow \pi^+\pi^-\gamma)_{IB}}\hskip 2pt ;\qquad
R_S=\frac{\Gamma(K_S\rightarrow \pi^+\pi^-\gamma)}{
\Gamma(K_S\rightarrow \pi^+\pi^-\gamma)_{IB}}\hskip 2pt .
\end{equation}
The factor $R_L$ contains all contributions to Eq.(\ref{digamma}). As already
mentioned, the IB contribution is suppressed by $\vert\eta_{+-}\vert^2$ and
it turns out that it is comparable to the magnetic contribution \cite{ramberg},
so that the numerical value of $R_L$ will be a factor $times$
${\vert\eta_{+-}\vert^2}$. Ideally, experiments should be able to fit all
the coefficients of the three time-dependent terms in (\ref{ievolutionkppg})
and (\ref{iievolutionkppg}), so that the corresponding interesting physics
could be measured. Using (\ref{hl}) and the
approximation $\eta_{+-}\cong \varepsilon$, we can write the leading
contribution to the interference as:
\begin{eqnarray}
I(\Delta t \stackrel{<}{\scriptstyle >}
 0 )_{\pi^+\pi^-}^{interf}&=&\frac{\Gamma(K_S\rightarrow \pi^+\pi^-)
\Gamma(K_S\rightarrow \pi^+\pi^-\gamma)_{IB}}
{16\gamma\vert p\vert^2\vert q\vert^2}\nonumber
\Big\{- 2\exp{(-\gamma\vert\Delta t\vert)}\vert\varepsilon \vert ^2
\nonumber\\
&\times&
\Big[\Bigl(1+\frac{\langle 2Re\hskip 2pt h_{E1}^S+
Re(\varepsilon^\prime_{\pi\pi\gamma}
h_{E1,ct}^{S}
/\varepsilon)\rangle_{int}}{\Gamma(K_S\to\pi^+\pi^-\gamma)_{IB}}\Bigr)
\cos{\Delta m \vert \Delta t\vert}\nonumber\\
&\mp&\frac{\langle Im(\varepsilon_{\pi\pi\gamma}^\prime h_{E1,ct}^{S}/
\varepsilon)
\rangle_{int}}{\Gamma(K_S\to\pi^+\pi^-\gamma)_{IB}}\sin{\Delta m
\vert\Delta t\vert}\Big]\Big\}.\label{phikppg}
\end{eqnarray}
Integrating over all times (essentially a few $\tau_S$), the intensity
resulting from direct $CP$ violation is of the order of
${\sim 10^{-9} \varepsilon_{\pi\pi\gamma}^\prime}$. Indeed,
direct $CP$ violation could be disentangled by considering
${I(\Delta t < 0 )_{\pi^+\pi^-}^{interf}-
I(\Delta t > 0 )_{\pi^+\pi^-}^{interf}}$. Although depressed by low
statistics (both at LEAR and at $\phi$-factories), these measurements have
the advantage over experiments at fixed-target beams \cite{barker} that the
different time behavior greatly helps in distinguishing the various terms.
Consequently, it might be not so surprising if better measurements or bounds
on ${\varepsilon^\prime_{\pi\pi\gamma}h_{E1,ct}^{S}
/\varepsilon}$ would come from
interferometry machines.

A difficulty is that all terms in (\ref{ievolutionkppg}) and
(\ref{iievolutionkppg}) are suppressed by at least
${\vert\varepsilon \vert ^2}$, and in addition there is the problem that
to distinguish them from each other requires the accurate knowledge of the
factors $R_L$ and $R_S$. In this regard, as an alternative to
$\pi^+\pi^-$ tagging one could use the semileptonic decay
${\pi^{\pm} l^{\mp} \bar{\nu} (\nu)}$. In this case one obtains:
\begin{eqnarray}
I(\Delta t < 0 )_{\scriptstyle \pi^{\pm} l^{\mp} \bar{\nu} (\nu)}
&=&\frac{\Gamma({\overline{K^0}}(K^0)\to \pi^{\pm} l^{\mp} \bar{\nu} (\nu))
\Gamma(K_S\rightarrow \pi^+\pi^-\gamma)_{IB}}
{16\gamma \vert \stackrel{\scriptstyle p}{\scriptstyle q}\vert ^2}\cr
&\times&\Big\{\exp{(-\Gamma_S\vert\Delta t\vert)} R_L
+\exp{(-\Gamma_L\vert\Delta t\vert)}R_S\cr
&\pm&2\exp{(-\gamma\vert\Delta t\vert)}
\Big[Re\hskip 2pt \Bigl(\eta_{+-}^*
+\frac{\langle h^{L*}_{E1}+\eta^*_{+-} h_{E1}^S\rangle_{int}}
{\Gamma(K_S\rightarrow \pi^+\pi^-\gamma)_{IB}}\Bigr)
\cos{\Delta m\vert\Delta t\vert}\nonumber \\
&+&Im \Bigl(\eta_{+-}^*+\frac{
\langle h^{L*}_{E1}+\eta^*_{+-} h_{E1}^S\rangle _{int}}
{\Gamma(K_S\rightarrow \pi^+\pi^-\gamma)_{IB}}\Bigr)
\sin{\Delta m\vert\Delta t\vert}\Big]
\Big\},\label{sievolutionkppg}
\end{eqnarray}
and
\begin{eqnarray}
I(\Delta t > 0 )_{\scriptstyle \pi^{\pm} l^{\mp} \bar{\nu} (\nu)}
&=&\frac{\Gamma({\overline{K^0}}(K^0)\to \pi^{\pm} l^{\mp} \bar{\nu} (\nu))
\Gamma(K_S\rightarrow \pi^+\pi^-\gamma)_{IB}}
{16\gamma \vert \stackrel{\scriptstyle p}{\scriptstyle q}\vert ^2}\cr
&\times&\Big\{\exp{(-\Gamma_L\Delta t)} R_L
+\exp{(-\Gamma_S\Delta t)}R_S\cr
&\pm&2\exp{(-\gamma\Delta t)}
\Big[Re\hskip 2pt \Bigl(\eta_{+-}^*+
\frac{\langle h^{L*}_{E1}+\eta^*_{+-} h_{E1}^S\rangle _{int}}
{\Gamma(K_S\rightarrow \pi^+\pi^-\gamma)_{IB}}\Bigr)
\cos{\Delta m\Delta t}\nonumber \\
&-&Im \Bigl(\eta_{+-}^*+\frac{\langle h^{L*}_{E1}+\eta^*_{+-} h_{E1}^S
\rangle_{int}}
{\Gamma(K_S\rightarrow \pi^+\pi^-\gamma)_{IB}}\Bigr)
\sin{\Delta m\Delta t}\Big]
\Big\}.\label{siievolutionkppg}
\end{eqnarray}
Here, the interference term is not suppressed by
${\vert\varepsilon\vert^2}$. Unfortunately, the cost is the
smaller ${Br(K_S\to\pi^{\pm} l^{\mp} \bar{\nu} (\nu))\sim 10^{-4}}$, instead of
${Br(K_S\rightarrow \pi^+\pi^-)}$ which appears in
(\ref{phikppg}). This gives a depressing factor
${Br(K_S\rightarrow\pi l\nu)\cdot Br(K_S\rightarrow\pi^+\pi^-\gamma)\sim
10^{-7}}$. Comparing (\ref{siievolutionkppg}) with (\ref{APP}), we see that
the interference terms have the same form, while (\ref{sievolutionkppg})
compared to (\ref{APP}) has just the opposite sign for the imaginary part.
Similar to the case of LEAR, we have here three different time-dependent
terms of the form (\ref{IBt}), (\ref{fdet}) and (\ref{cpt}). However, the
$\phi$-factory has the advantages that {\it i}): in principle one can
select the interference term by considering the asymmetry between opposite
charge modes,
${A(\Delta t\stackrel{<}{\scriptstyle >}0)
\equiv I(\Delta t \stackrel{<}{\scriptstyle >}0
)_{\scriptstyle \pi^{+} l^{-} \bar{\nu} }-
I(\Delta t\stackrel{<}{\scriptstyle >}0)_{\scriptstyle \pi^{-} l^{+}\nu}}$;
and {\it ii}): the imaginary part of the interference can be separately
studied by considering the difference ${A(\Delta t <0)-A(\Delta t >0)}$.
Concerning statistics, taking into account the suppression factor mentioned
above, it appears that, with {\it e.g.} ${10^{12}}$ $\phi'$s one might
obtain about 100 events related to the term (\ref{IBt}), and (at least) put
some significant constraints on the other two terms. In any case, this
analysis should enable to give limits on
${\varepsilon_{\pi \pi \gamma}^\prime}$ in a way
complementary to the direct measurement of the charge asymmetry
in ${K^\pm\rightarrow \pi^\pm\pi^0 \gamma}$ \cite{riaz}.

Another interesting issue to be pursued at the $\phi$-factory is the intensity
with the $\theta$-cut Dalitz plot defined in (\ref{theta}), giving access to
$h_{E2}^L$. For ${f_1=\pi^+\pi^-}$ tagging, this can be expressed as:
\newpage
\begin{eqnarray}
I(\Delta t < 0)^\theta_{\scriptstyle \pi^+\pi^-}
&=&\frac{\Gamma(K_S\rightarrow \pi^+\pi^-)}
{16\gamma\vert p\vert^2\vert q\vert^2}
\big\{2\langle Re(h^L_{E2}\varepsilon^*)\rangle_{int}^{\theta}
\exp{(-\Gamma_{S}\vert\Delta t\vert)}
- 2\exp{(-\gamma\vert\Delta t\vert)}\nonumber\\
&&\times\big[\langle Re(\varepsilon^* h_{E2}^L)\cos{\Delta m\vert\Delta t\vert}
-Im(\varepsilon^*h_{E2}^L)\sin{\Delta m
\vert \Delta t\vert}\rangle_{int}^\theta\big]\big\},\label{phitetakppg1}
\end{eqnarray}
and
\begin{eqnarray}
I(\Delta t > 0)^\theta_{\scriptstyle \pi^+\pi^-}
&=&\frac{\Gamma(K_S\rightarrow \pi^+\pi^-)}
{16\gamma\vert p\vert^2\vert q\vert^2}
\big\{2\langle Re(h^L_{E2}\varepsilon^*)\rangle_{int}^{\theta}
\exp{(-\Gamma_{L}\Delta t)}
- 2\exp{(-\gamma\Delta t)}\nonumber\\
&&\times\big[\langle Re(\varepsilon^* h_{E2}^L)\cos{\Delta m\Delta t}
+Im(\varepsilon^*h_{E2}^L)\sin{\Delta m
\Delta t}\rangle_{int}^\theta\big]\big\}.\label{phitetakppg2}
\end{eqnarray}

Analogously, for semileptonic tagging:
\begin{eqnarray}
I(\Delta t < 0)^\theta_{\scriptstyle\pi^{\pm}l^{\mp}\bar{\nu}(\nu)}
&=&\frac{\Gamma({\overline{K^0}}(K^0)\to\pi^{\pm} l^{\mp}\bar{\nu}(\nu))}
{16\gamma \vert\stackrel{\scriptstyle p}{\scriptstyle q}\vert^2}
\big\{2\langle Re(h^L_{E2}\varepsilon^*)\rangle_{int}^{\theta}
\exp{(-\Gamma_{S}\vert \Delta t\vert)}\nonumber\\
&&\pm2\exp{(-\gamma\vert\Delta t\vert)}\hskip 2pt \big[\langle
Re\hskip 2pt h_{E2}^L
\cos{\Delta m\vert\Delta t\vert}
-Im\hskip 2pt h_{E2}^L
\sin{\Delta m\vert\Delta t\vert}\rangle_{int}^\theta\big]\big\},
\label{phithetakpilnu1}\end{eqnarray}
and
\begin{eqnarray}
I(\Delta t > 0)^\theta_{\scriptstyle\pi^{\pm}l^{\mp}\bar{\nu}(\nu)}
&=&\frac{\Gamma(K^0(\bar{K^0})\rightarrow \pi^{\pm} l^{\mp}\bar{\nu}(\nu))}
{16\gamma\vert\stackrel{\scriptstyle p}{\scriptstyle q}\vert^2}
\big\{2\langle Re(h^L_{E2}\varepsilon^*)\rangle_{int}^{\theta}\exp{(-\Gamma_{L}
\Delta t)}\nonumber\\
&&\pm 2\exp{(-\gamma\Delta t)}\hskip 2pt
\big[\langle Re\hskip 2pt h_{E2}^L\cos{\Delta m\Delta t}
+Im\hskip 2pt h_{E2}^L\sin{\Delta m
\Delta t}\rangle_{int}^\theta\big]\big\}.\label{phithetakpilnu2}
\end{eqnarray}

We notice that phenomenologically ${(h^L_{E2}\varepsilon^*)}$ tends to be
almost purely imaginary in the framework considered above, {\it i.e.}
ChPT to order $p^4$, since
${\arg{\varepsilon}\sim 44^\circ}$ and, from the definition (\ref{fde}),
final state interactions determine ${\arg{h^L_{E2}}\sim-\delta_0}$
with $\delta_0$ is the $I=l=0$ $\pi\pi$ phase shift. It turns
out that this angle is about $(39\pm 5)^\circ$ \cite{ochs}, which implies
${\arg{h^L_{E2}\varepsilon^*}\sim -83^\circ}$. Consequently, only the last
term in (\ref{phitetakppg1}) and (\ref{phitetakppg2}) is substantially
different from zero. Instead, with the semileptonic tagging, both oscillating
terms in (\ref{phithetakpilnu1}) and (\ref{phithetakpilnu2}) are different
from zero, while also in this case the purely exponential term tends to
be almost vanishing. Anyway, we can more precisely dispose of this term and
select the oscillating interference by considering the difference between
intensities with opposite lepton charges in the tagging channel. Furthermore,
$\sin\Delta m \Delta t$ can be isolated by the time asymmetry. Concerning
the needed statistics,  at least ${10^{12}}\ \phi$'s should be required
for this kind of analysis.
\section{Conclusions}
\label{sec:concl}
In this paper we have analyzed the advantages of using interferometry kaon
machines like LEAR or the $\phi$-factory in the study of the decays
$K_{L,S}\rightarrow 3\pi$ and $K_{L,S}\rightarrow\pi\pi\gamma$. Typical
interference patterns can be obtained by studying as a function of time the
difference ${\Gamma(K^0\to f)-\Gamma({\overline{K^0}}\to f)}$
at LEAR, and at the $\phi$-factory the intensity for $\phi$-decaying to the
$K_S K_L$ system and this in turn to two final states $f_1$ and $f_2$.
Generally, at the $\phi$-factory one of the final states is chosen to be a
tagging channel \cite{fukawa}, and we have considered here both two pions and
semileptonic decays tagging.

Concerning the channel $K\to 3\pi$, we have seen how LEAR could measure the
$(3\pi)$ phase shifts, by just fitting the correlation
${A(K_L\to 3\pi)^*A(K_S\to 3\pi)}$ to the data as a function of time.
Actually, it seems
already possible to put some interesting limit on these phases using the
present data. This analysis extends the previous discussion of this issue,
presented for the case of the $\phi$-factory in Ref.\cite{dambrosio1}.

As to the channel $K\to\pi\pi\gamma$, the correlation
${A(K_L\to\pi^+\pi^-\gamma)^*A(K_S\to\pi^+\pi^- \gamma)}$ has a
richer structure, which we have tried to analyze for both the $\phi$-factory
and LEAR. This correlation can be either symmetric or antisymmetric under
exchange of pion four-momenta. Accordingly, to select the corresponding
physics, we suggest to analyze the data by integrating either over the full
Dalitz plot or with an antisymmetric kinematical cut. In the symmetric
correlation, several interesting physical effects can be either studied
or significantly constrained. These include: the $CP$ violation in
$A(K_L\rightarrow \pi^+ \pi^-\gamma)_{IB}$, proportional to
$A(K_L\rightarrow \pi^+ \pi^-)$; the $CP$ conserving structure dependent
amplitude $A(K_S\to\pi^+\pi^-\gamma)_{DE}$; the direct $CP$ violation
$\varepsilon^\prime_{\pi\pi\gamma}$, in general not suppressed
by the $\Delta I=1/2$ rule. Our discussion indicates that interferometry
machines should be quite useful to study these effects, and that
time-dependence in correlations provides a convenient way to disentangle
direct $CP$ violation from other contributions. Furthermore, in the case of
the $\phi$-factory with semileptonic tagging, one can define time asymmetries
which are directly proportional to $\varepsilon^\prime_{\pi\pi\gamma}$.
Also, intensities integrated asymmetrically over the phase space both at
LEAR and at the $\phi$-factory should be an efficient tool to measure
$CP$ conserving higher multipole amplitudes.

A related analysis was performed in Ref.\cite{donoghue}, where quantum
correlations in $K\to\pi^+\pi^-\gamma$ were studied, limiting to the
$\phi$-factory with two pion tagging
and (mostly) to large time intervals ${\Delta t\to\infty}$,
where interference is not so important. Here, we complement that analysis in
several directions, namely we consider also the case of LEAR and include
higher multipole amplitudes in the analysis. In addition, for the
$\phi$-factory, we are mostly concerned with finite time intervals
$\Delta t$, for which interference plays a crucial role. Furthermore, also
the semileptonic tagging has been exploited in the present paper.

In conclusion, we expect that very likely, by time-dependence measurements,
interferometry machines will
measure the three-pion phase shifts, will improve the
existing value of the $CP$ conserving $K\to\pi\pi\gamma$ amplitude and
put a stringent limit on $\varepsilon^\prime_{\pi\pi\gamma}$.
These time-dependence measurements should usefully complement higher
statistics experiments at fixed-target kaon beams.

\end{document}